\documentclass[11pt,a4paper]{article}
\pdfoutput=1

\usepackage{bmpsize}
\usepackage{jheppub}
\usepackage{graphicx,bm,color,wrapfig,cancel}
\usepackage{amsmath,amssymb,bm}
\usepackage{float}
\usepackage{braket}
\usepackage{comment}
\usepackage{subfigure}
\usepackage[utf8]{inputenc}
\usepackage{appendix}
\usepackage{ulem}

\newcommand{\fulltoday}{\number\day\space \ifcase\month\or
    January\or February\or March\or April\or May\or June\or
    July\or August\or September\or October\or November\or December\fi
    \space\number\year}

 \makeatletter
    
    \@addtoreset{equation}{section}
  \makeatother

\allowdisplaybreaks[3]

\title{\boldmath
Detecting scale anomaly in chiral phase transition of QCD: new critical endpoint pinned down 
}

\author[a,b]{Mamiya Kawaguchi,}
\author[c]{Shinya Matsuzaki,}
\author[d,e]{Akio Tomiya}

\affiliation[a]{Department of Physics and Center for Field Theory and Particle Physics, Fudan University, 220 Handan Road, 200433 Shanghai, China}

\affiliation[b]{
School of Nuclear Science and Technology, University of Chinese Academy of Sciences, Beijing 100049, China}

\affiliation[c]{Center for Theoretical Physics and College of Physics, Jilin University, Changchun, 130012, China}

\affiliation[d]{RIKEN BNL Research center, Brookhaven National Laboratory, Upton, NY, 11973, USA}

\affiliation[e]{Department of Information Technology, International Professional University of Technology in Osaka, 3-3-1 Umeda, Kita-Ku, Osaka, 530-0001, Japan }

\emailAdd{kawaguchi@fudan.edu.cn}
\emailAdd{synya@jlu.edu.cn}
\emailAdd{akio.tomiya@riken.jp}

\abstract{
Violation of scale symmetry, scale anomaly, being a radical concept in quantum field theory, is of importance to comprehend the vacuum structure of QCD, and should potentially contribute to 
the chiral phase transition in thermal QCD, 
as well as the chiral and U(1) axial symmetry.  
Though it should be essential, 
direct evidence of scale anomalies has never been 
observed in the chiral phase transition.  
We propose a methodology to detect 
a scale anomaly in 
the chiral phase transition, which is an electromagnetically induced scale anomaly:   
apply a weak magnetic field background onto 
two-flavor massless 
QCD with an extremely heavy strange quark, 
first observe the chiral crossover;  
second, 
adjusting the strange quark mass to be smaller and smaller, 
observe the second-order chiral phase transition, and 
then the first-order one in the massless-three flavor limit. 
Thus, the second-order chiral phase transition, observed as 
the evidence of the quantum scale anomaly, is 
a new critical endpoint.
It turns out that this electromagnetic scale anomaly gets  
most operative in the weak magnetic field regime, rather than 
a strong field region. 
We also briefly address accessibility of lattice QCD, 
a prospected application to dense matter system, 
and implications to astrophysical observations, such 
as gravitational wave productions provided from 
thermomagnetic QCD-like theories.   
}

\begin{document}





\maketitle
\flushbottom
%

%


\section{Introduction}

The origin of scales  
in QCD 
can be classified into 
two categories in terms of quantum field theory: 
one is spontaneous symmetry breaking of chiral symmetry, 
while the other  
explicit symmetry breaking. 
The latter 
is subject to renormalization of the quantum corrections in QCD, quantum scale 
anomalies, and the presence of quark masses (classical- chiral and scale breaking). 
Applying QCD in a thermal bath would have a potential  
to detect the direct consequence of the spontaneous 
breaking of the chiral symmetry, by observing 
the damping of the order parameter for the chiral 
phase transition (i.e. the quark condensate) at high temperatures above 
the critical temperature. 
Thus, the scale breaking associated with the 
spontaneous breaking can be dropped out there.   
However, there is still inevitable contamination left:  
that is the scale anomaly arising from renormalization evolution of the QCD gauge coupling, 
which should be present all the way in the thermal history of QCD, 
and it still coexists with the scale breaking from quark masses, 
even above the critical temperature. 
Hence detection of the scale anomaly   
seems to be quite challenging even in the thermal QCD.



\begin{figure}[!htpb]
  \begin{center}
   \includegraphics[width=10cm]{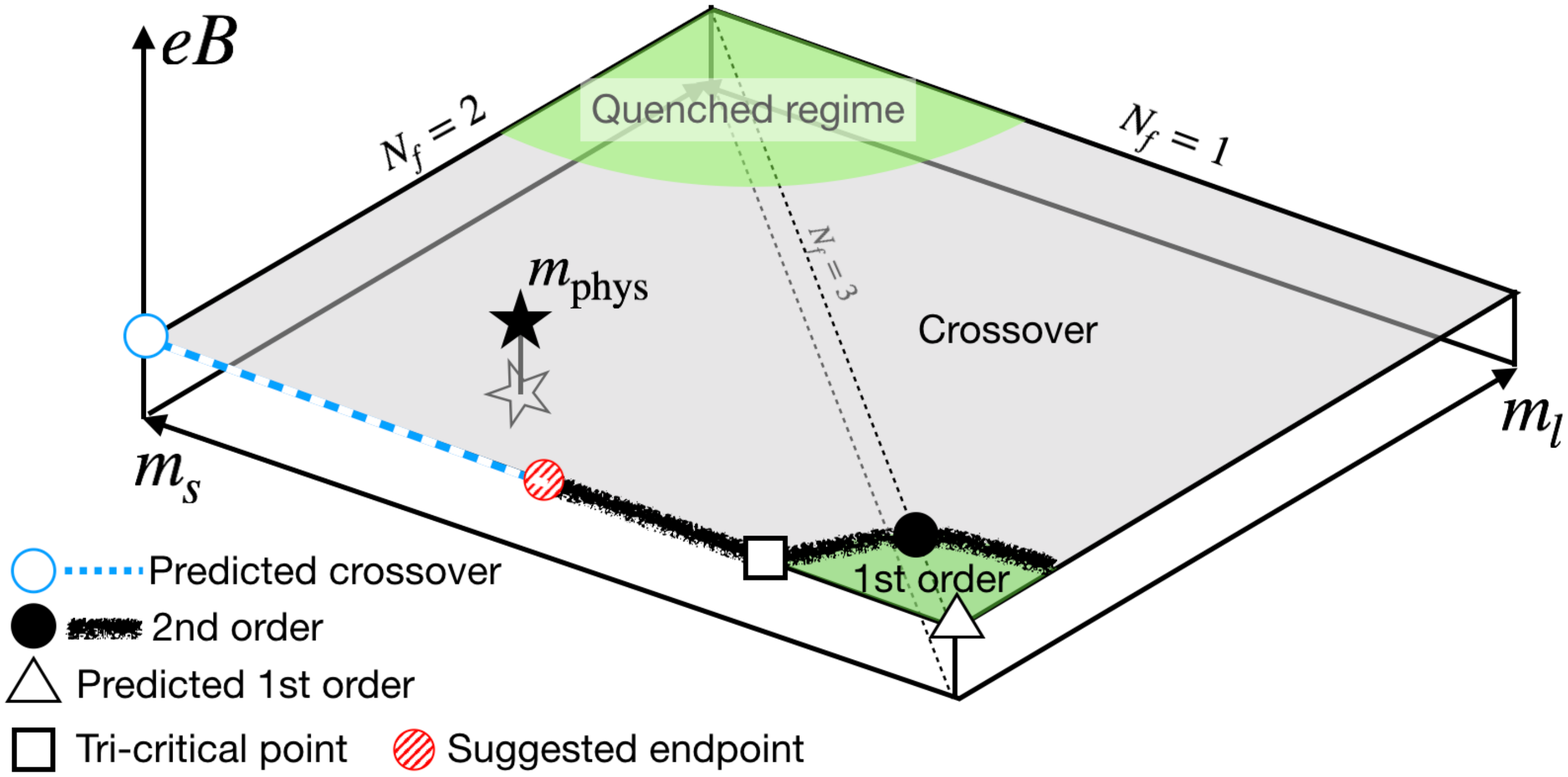}
  \end{center}   
\vspace{-0.5cm}
\caption{
Extended Columbia plot along with an external magnetic field in the weak field regime.
$m_l$ and $m_s$ denote masses of the 
light quarks and strange quark, respectively.
The existence of the scale anomaly is dictated as 
the suggested endpoint on this Columbia plot, 
indicated by a shaded circle. 
A blank square stands for the tri-critical point, at which 
the three domains of 
crossover, first- and second-order phase transitions  
merge. 
The details are described in the text. 
The physical point is known to be the crossover \cite{Endrodi} at this size of magnetic field.
\label{fig:columbia_plot}
}  
\end{figure}



To reconcile this dilemma, it would be necessary to  
go away, from the physical point, at the massless limit (so-called the chiral limit). 
There quarks enjoy the exact chiral symmetry and the scale 
symmetry breaking is provided purely by the quantum scale anomaly. 
However, the dimensional transmutation 
in the infrared energy regime, associated with the scale anomaly,  
triggers the spontaneous chiral breaking, hence contaminates with each other 
in the thermal evolution of the chiral order parameter. 
Thus the case would still be involved even in the chiral limit. 

Pisarski and Wilczek a long time ago \cite{Pisarski-Wilczek} payed their 
particular attention to the chiral limit, off the physical point, and employed 
an effective model in the confinement phase based on the exact 
chiral symmetry with or without $U(1)$ axial symmetry. 
They discussed the renormalization group runnings of 
the model couplings, and 
existence of the renormalization 
group-fixed point. 
It was suggested that the order of the chiral 
phase transition highly depends on 
the number of massless flavors (namely the chiral symmetry 
and $U(1)$ axial symmetry). 
This was the pioneering work and made a significant help 
to comprehend QCD at the physical point as the chiral theory. 
However, at this point, it is yet unclear how the scale anomaly 
directly affects the chiral phase transition.

Columbia university group extended their work using lattice QCD calculations \cite{Columbia}.
In lattice QCD, mass and the number of flavor dependence on the order of the QCD phase transition 
have been investigated 
including the chiral 
and confinement-deconfinement phase transitions. 
It is summarized as what is called Columbia plot, 
which is drawn on the quark mass plane, hence 
dictates the chiral and $U(1)$ axial symmetry structures. 

Since the early work of Columbia group used an approximated algorithm, 
it has nowadays been refined by utilizing modern technologies. 
\textcolor{black}{
In the two flavor chiral limit and
2+1 flavor chiral limit, namely with two of massless quarks and one heavy quark, 
restoration or persistence of $\mathrm{U(1)}$ axial anomaly above the critical temperature has been discussed
\cite{JLQCD-overlap, JLQCD-DW, JLQCD-DW2, HotQCD-DW1, HotQCD-DW2, HotQCD-DW3, Sheng-Tai, HISQ-2+1}. 
It is shown that restoration or persistence is tightly connected to the order of phase transition and its universality class \cite{Pisarski-Wilczek, Aoki-Fukaya-Taniguchi, u1RG, Bootstrap}.
On the other hand, QCD is believed to exhibit the first-order phase-transition in the three-flavor chiral limit \cite{Pisarski-Wilczek}. 
However, no clear lattice results for the first-order phase-transition has been obtained and only lower limits on the quark mass in the crossover 
has been placed~\cite{3F-HISQ, 3F-Wilson}.
}

Thus the chiral properties in the chiral phase transition 
have extensively been studied 
on lattice QCD and so far been well understood through projection on to 
the Columbia plot, initiated from the Pisarski-Wilczek's work.  
However, never have the scale anomaly  been clearly understood 
or given any definite signal 
even on the Columbia plot:   
the form of the QCD scale anomaly is flavor universal, hence would survive all the way 
on the Columbia plot.
Thus it may be conclusive that the QCD scale anomaly leaves no distinct footprint on the Columbia plot.

In this paper, we propose a methodology to detect 
another scale anomaly in 
the chiral phase transition on a Columbia plot:   
apply a weak magnetic field background onto QCD, 
and see that an electromagnetically induced scale anomaly arises, which is 
coupled to the chiral order parameter; 
work in two-flavor massless 
QCD with an extremely heavy strange quark;    
first observe the chiral crossover; second, 
adjusting the strange quark mass to be smaller and smaller, 
observes the second-order chiral phase transition, 
and the first-order one in the massless-three flavor limit. 
See Fig.~\ref{fig:columbia_plot}.

That is an evidence of 
the quantum scale anomaly (an electromagnetically induced 
quantum scale anomaly), which 
turns to 
transparently contribute to the chiral phase transition, 
so is the essential key to significantly affect 
the Columbia plot, 
as well as the chiral and $U(1)$ axial symmetries.  
It turns out that this electromagnetic scale anomaly gets  
most operative in the weak magnetic field regime. 
In the massless-three flavor limit, this quantum scale 
anomaly contribution is overwhelmed by 
the U(1) axial anomaly's, 
hence  
the first-order phase transition is realized, while in the massless-two flavor limit such interference does not happen, and the crossover is promoted by the induced tadpole,   
as will explicitly 
be demonstrated later. 
Thus, the second-order chiral phase transition, observed as 
the evidence of the quantum scale anomaly, is 
a new critical endpoint. 

Other critical points have been proposed in a strong magnetic field regime~\cite{Cohen:2013zja,Endrodi:2015oba}. Our finding is definitely different from those which  are expected to emerge due to broadening of the quenched regime in Fig.~\ref{fig:columbia_plot} because of the amplified mass gap of quarks by the strong magnetic field.

Lots of interesting results have been reported from lattice studies on the QCD thermodynamics in the strong external magnetic field \cite{Dima-top, Massimo, Dima-vortex, Endrodi-butterfly}.  
Of these results, there are two striking properties discovered: the reduction of transition temperature \cite{ catalysis2, endrodi-eos, Tomiya-Hisq, Massimo-heavy-pion, Enrodi-heavy-pion}
and the inverse magnetic catalysis \cite{Endrodi}. 
In contrast to the strong magnetic field 
the weak magnetic field has not fully been investigated on the lattice QCD at non-zero temperature. 
At the current status, 
the smallest nonzero magnetic field on the lattice is  
%
$\sqrt{eB}_\text{min}\sim 0.1$ GeV~\cite{Endrodi},  
which is only for the physical pion mass simulations at finite temperatures~\footnote{On the lattice, the minimum of magnetic field is bounded by the temperature,
$|eB|_\text{min} \propto T^2$.
See~\cite{Endrodi} for details.}. 
This is because of the fact that 
the minimal magnetic field is determined by the area of transverse plane to the magnetic field~\cite{Endrodi}.
Thus the weak magnetic field 
with smaller pion masses, particularly close to the chiral limit, is challenging due to the higher numerical cost.

Instead of lattices, 
chiral effective theories, like what we will employ, 
would therefore shed light 
on the weak magnetic regime at the chiral limit,  
as if it gets back to the epoch where the Columbia plot was first proposed along the pioneering work by Pisarski and Wilczek: our finding 
paves a way to investigate this new frontier as searching for the evidence of the quantum scale anomaly in the chiral phase transition.

The proposed critical endpoint is marked 
as ``Suggested endpoint", in 
an extended Columbia plot including the magnetic field in the weak field regime,  Fig.~\ref{fig:columbia_plot}. 
The new endpoint reveals the potential existence of the 
scale anomaly in QCD, which is made transparent in the presence of 
a magnetic field. 
This is our main result in the present paper.

\section{Ginzburg-Landau description}

Before entering the detailed demonstration, we shall present 
an intuitive interpretation on the chiral phase transition nature,  
which would help readers to easily grasp 
our new finding. 
To this end, we momentarily work on a generic Ginzburg-Landau 
description for the chiral phase transition, 
inspired by 
Pisarski and Wilczek~\footnote{
Our current analysis is based on an effective potential approach, 
not on the renormalization group method like in the 
Pisarski and Wilczek's work. 
Cross-check using the (nonperturbative) renormalization group 
would make our finding more evident, which will be pursued elsewhere. 
}.

In the Ginzburg-Landau approach,
the chiral phase structure can be described by
a generic effective potential 
in terms of the order parameter $\bar \sigma_0$. 
It takes the polynomial form.  
\begin{eqnarray}
V_{\rm eff}(\bar\sigma_0)&=&
\alpha_1(T,eB)\bar \sigma_0+
\alpha_2(T,eB)\bar \sigma_0^2
\nonumber\\
&&
+
\alpha_3(T,eB)\bar \sigma_0^3+
\alpha_4(T,eB)\bar \sigma_0^4+
\cdots 
\label{GL-model}
\end{eqnarray}
where the potential parameters $\alpha_{i=1,2,3,4,\cdots}$ can include intrinsic temperature- and/or  magnetic-dependence.
Our sigma field is defined as an interpolating mesonic 
degree of freedom for the lightest quark bilinear $\bar{q}q$ in QCD.

When the exact chiral symmetry is imposed, the effective potential form is 
restricted to be an even function of 
$\bar \sigma_0$, such as $\bar \sigma_0^2$ and $\bar \sigma_0^4$. 
With terms truncated up to $O(\bar \sigma_0^4)$, 
the phase transition is then expected to be of the second order.

In the three-flavor case,
the $U(1)$ axial anomaly induces the cubic term of $\bar \sigma_0$, which originally takes 
a determinant form of the quark condensate, ${\rm det} \bar{q}q \sim \bar{\sigma}_0^3$, 
a la Kobayashi-Maskawa-'t Hooft~\cite{Kobayashi:1970ji,Kobayashi:1971qz,tHooft:1976rip,tHooft:1976snw}. 
Then, the phase transition order is changed to be of the first order, 
because the cubic term creates a potential barrier between 
the chiral symmetric vacuum $(\bar{\sigma}_0 =0)$ and broken one 
$(\bar{\sigma}_0 \neq 0)$. 
Note here that the determinant term in the two-flavor case can 
be absorbed into $\bar{\sigma}_0^2$ term, 
so the $U(1)$ axial anomaly does not 
effectively affect the potential form in Eq.(\ref{GL-model}).

As to the linear (tadpole) term of $\bar{\sigma}_0$, 
it cannot be present even with the $U(1)$ axial anomaly. 
It would show up when the current quark masses are introduced, 
which can explicitly break the full chiral symmetry including the 
$U(1)$ axial part.  
Once the tadpole is present, the potential no longer 
achieves the exactly chiral symmetric vacuum, which, in other words, 
the theory always lies in the broken phase, even though 
the value of $\bar{\sigma}_0$ gets smaller, as $T$ becomes higher. 
This implies that the tadpole term in the potential tends to drive  
the phase transition to be continuous, i.e., crossover, called 
the chiral crossover.


To make the point better understood, 
we may explicitly introduce the chiral field $\Phi$, 
which is parametrized by the scalar- and pseudoscalar-meson fields as 
\begin{eqnarray} 
\Phi=
\begin{cases}
\frac{1}{2}(\sigma_0 +i\tau^i\pi^i ),\;\;\;
(\hbox{for }N_f=2)\\
\frac{1}{2}(\sigma_a+i\pi_a)\lambda^a,\;\;\;
(\hbox{for }N_f=3), 
\end{cases}
\end{eqnarray}
where
$\tau_{i=1,\cdots,3}$ are the Pauli matrices
and 
$\lambda_{a=1,\cdots,8}$ represent the Gell-Mann matrices
with $\lambda_0=\sqrt{{2/3}} \cdot {\bm 1}_{3\times 3}$. 
Under the chiral symmetry  
we define the transformation law of $\Phi$ as 
\begin{eqnarray}
\Phi\to 
\begin{cases}
g_L \cdot \Phi \cdot g_R^\dagger,\;\;\;(\hbox{for }N_f=2) \\
g_A\cdot g_L \cdot \Phi \cdot g_R^\dagger
,\;\;\;(\hbox{for }N_f=3),
\end{cases}
\label{chiralT}
\end{eqnarray}
where  
$g_{L,R}\in SU(2)_{L,R}$
in the case of
$N_f=2$,
and 
$g_{L,R}\in SU(3)_{L,R}$ with $g_A\in U(1)_A$
in the case of $N_f=3$. 

Thus the linear sigma model as the Ginzburg-Landau description is constructed 
based on the chiral invariance with its breaking 
source, including operators up to dimension four, as 
\begin{eqnarray}
{\cal L}_{\rm LSM}&=&
{\rm tr}\left[ \partial_\mu \Phi\partial^\mu \Phi^\dagger \right]
-V(\Phi), 
\label{Lag:LSM}
\end{eqnarray}
where $V(\Phi)$ represents the potential terms,
\begin{eqnarray}
V(\Phi)&=&
V_0(\Phi)+V_{\rm anom}(\Phi)+V_{\rm SB}(\Phi).
\label{VPhi}
\end{eqnarray}

In Eq.(\ref{VPhi}) $V_0(\Phi)$ is the chiral invariant part: 
\begin{eqnarray}
V_0(\Phi)=\mu^2{\rm tr }[(\Phi^\dagger\Phi)]
+\lambda_1{\rm tr }[(\Phi^\dagger\Phi)^2]
+\lambda_2({\rm tr }[(\Phi^\dagger\Phi)])^2, 
\end{eqnarray}
where 
$\mu^2$ is the mass parameter taken either a positive or negative value. 
$\lambda_{1,2}$ are dimensionless quartic coupling constants, which cannot be taken independently  
in the case of $N_f=2$, because of the speciality of 
$SU(2)$ algebra.

For $N_f=3$, the $U(1)$ axial anomalous part, but keeping the chiral $SU(N_f)_L\times SU(N_f)_R$ symmetry, represents $V_{\rm anom}(\Phi)$.
The lowest dimensional operator for the $U(1)$ axial anomalous part
is given by a la Kobayashi-Maskawa-'t Hooft~\cite{Kobayashi:1970ji,Kobayashi:1971qz,tHooft:1976rip,tHooft:1976snw},
\begin{eqnarray}
V_{\rm anom}(\Phi)
=
-B_{\rm anom}\left({\rm det}[\Phi]+{\rm det}[\Phi^\dagger]\right),
\label{Vanom}
\end{eqnarray}
where
the parameter $B_{\rm anom}$ is taken to be real, 
which has mass dimension one.

In the case of $N_f=2$, $V_{\rm anom}$ is redundant: 
${\rm det} \Phi +{\rm h.c.}= 2 (\sigma_0^2 + (\pi_i)^2)$
hence is indistinguishable from 
the $\mu^2$ term, because $\Phi$ has no $U(1)_A$ charge as in Eq.(\ref{chiralT}).

In the underlying QCD Lagrangian, 
the chiral symmetry is explicitly broken by 
 the current quark mass matrix ${\cal M}={\rm diag}\{m_u,m_d (,m_s) \}$. 
In the linear-sigma model-Lagrangian,
this explicit breaking effect is reflected in 
the $V_{\rm SB}$ part: 
\begin{eqnarray}
V_{\rm SB}(\Phi)=
-c \, {\rm tr}[{\cal M}\Phi^\dagger+{\cal M}^\dagger \Phi],
\end{eqnarray}
where 
the parameter $c$ is taken to be a real value, 
and has mass dimension two.


To be consistent with the underlying QCD, 
we choose the background-field profile for the $\Phi$ field 
to respect the vectorial symmetry, $SU(2)_V$ or $SU(3)_V$, 
in such a way that 
\begin{eqnarray}
\bar \Phi \cdot {\bf 1}=
\begin{cases}
\frac{1}{2}\bar \sigma_0 \cdot {\bf 1}_{2 \times 2},
\;\;\;\;\;\;
(\hbox{for }N_f=2)
\\
\frac{1}{2}\sqrt{\frac{2}{3}}
\bar \sigma_0 \cdot {\bf 1}_{3 \times 3},
\;\;\;
(\hbox{for }N_f=3).
\end{cases}
\end{eqnarray}
Thus the linear sigma model potential can be viewed as 
a Ginzburg-Landau description,     
at the tree level (i.e. mean field level): 
\begin{eqnarray}
V_{\rm tree}(\bar \sigma_0)
=
\begin{cases}
-\sum_f cm_f\bar\sigma_0
+\frac{1}{2}(\mu^2-B_{\rm anom})\bar \sigma_0^2
+(\frac{\lambda_1}{8}
+\frac{\lambda_2}{4})\bar \sigma_0^4
,\;\;\;
({\rm for}\;\;\;N_f=2 )\\
-\sqrt{\frac{2}{3}}\sum_f cm_f\bar\sigma_0
+\frac{1}{2}\mu^2\bar \sigma_0^2
-\frac{B_{\rm anom}}{3\sqrt{6}}\bar \sigma_0^3
+(\frac{\lambda_1}{12}
+\frac{\lambda_2}{4})\bar \sigma_0^4,\;\;\;
({\rm for}\;\;\;N_f=3 ).
\end{cases}
\label{tree_pot}
\end{eqnarray}

From this potential,
the vacuum 
expectation value $\langle \bar \sigma_0 \rangle $ can be determined from the stationary condition,
\begin{eqnarray}
\frac{\partial V_{\rm eff}(\bar \sigma_0)}{\partial \bar \sigma_0} 
\Bigg|_{\bar{\sigma}_0 = \langle  \bar \sigma_0 \rangle 
= \sqrt{ 2 N_f } \langle  \bar{\Phi} \rangle} 
=0.
\label{sc}
\end{eqnarray}
At the tree level evaluation, 
it is related to the pion decay constant $f_\pi$ as 
\begin{eqnarray}
\langle \bar{\Phi} \rangle  
=
\frac{1}{\sqrt{2 N_f}} \langle \bar{\sigma}_0 \rangle  
=f_\pi/2 .
\end{eqnarray}

We can also discuss the order of chiral phase transition from the tree-level potential in Eq.~(\ref{tree_pot}).
At the chiral limit for the case of $N_f=2$ ($m_f=0$), 
the parameter $B_{\rm anom}$ can be absorbed in
the mass parameter $\mu^2$, so that 
the $U(1)$ axial anomaly part does not effectively affect the 
the order of chiral phase transition. The tree-level potential then has only the $\bar \sigma_0^2$ and $\bar \sigma_0^4$ terms. 
Thus, in the massless two-flavor case,
the phase transition is deduced to be of the second order.  

In the massless three-flavor case,
the $U(1)$ axial anomaly part generates the cubic term of $\bar \sigma_0$, which 
creates a potential barrier between the chiral symmetric vacuum ($\bar \sigma_0=0$) and broken one ($\bar \sigma_0\neq 0$).
In association with the deformation of the potential,
the phase transition is changed to be of the first order.

We will later observe that a tadpole term is generated 
by the electromagnetic scale anomaly, even for
the massless flavor cases, in the thermomagnetic QCD. 
The phase transition becomes crossover, so 
there should be a critical endpoint seen in the interplay between 
massless two-flavor and three-flavor thermomagnetic QCD, as in Fig.~\ref{fig:columbia_plot}.



\section{Scale anomaly coupled to chiral order parameter}

In this section we discuss how the chiral order parameter 
can affect the scale anomaly. 
First, note that in the low-energy meson-dynamics, the scalar meson fields serve as a source for the scale symmetry breaking in QCD, 
accompanied with the chiral symmetry breaking. 
Then the scale (dilatation) current $j_D(x)$ is composed of hadrons, and 
would couple to the chiral singlet/isosinglet states,  
involving 
the (chiral singlet/isosinglet component of) $\sigma$ meson state, the two-pion resonant state, 
a tetraquark state, and glueball, and so forth.  
Since we work on the chiral phase transition 
by the Ginzburg-Landau description 
and are interested particularly in a coupling of electromagnetic field background to the chiral order 
parameter, 
only the chiral singlet/isosinglet component of the meson field  
is relevant.

We also assume the lightest isoscalar mesons to be mostly 
composed of quarkonium state, so that the vacuum expectation value associated with the tetraquark state is negligible.  
One might think that a glueball field can mix with the 
sigma meson field, hence might contribute to the scale anomaly 
relevant to the chiral phase transition. 
However, it has recently been reported from lattice simulations for 
$2 + 1$ flavors that 
around the chiral limit which is currently our main concern, 
the fluctuation of Polyakov loop, which can be regarded 
as (an electric part) of glueball, does not have significant correlation with   
the chiral order parameter~\cite{Clarke:2019tzf}. 
This indicates negligible mixing between the glueball  
and sigma meson around the chiral limit (with mass less than 
the physical-point value). 
Hence such a gluonic term will not be taken into account 
in the present work. 
%

Thus, as far as the coupling to the chiral order parameter at the low-energy is concerned, 
we can approximate the overlap amplitude between the scale current 
and (the chiral singlet/isosinglet component of) 
the sigma meson state, dubbed $\phi$, 
as~\footnote{
In Eq.~(\ref{PCDC}),
the 
overall sign can be minus, and then it can be absorbed in the decay constant $f_\phi$. 
} 
\begin{eqnarray}
\langle 0|
\partial_\mu j^\mu_D(x)| \phi(p)   \rangle
=
\langle 0|
\left(T^\mu_{\;\;\mu}\right)_{\rm meson}(x)| \phi(p)   \rangle 
\approx 
f_\phi m_\phi^2
e^{-ip\cdot x}
, 
\label{PCDC}
\end{eqnarray}
where we have taken 
$\langle \phi \rangle = f_\phi$, which plays the role of 
the decay constant for the scale breaking, and 
$\left(T^\mu_{\;\;\mu}\right)_{\rm meson}$ is
the trace of the energy momentum tensor described by mesons.
%


\subsection{Ward-Takahashi identity for scale symmetry in electromagnetic field}

The scale symmetry is also explicitly broken by the electromagnetic contribution and becomes anomalous.
As a consequence, 
the chiral-singlet/isosinglet scalar can be coupled to 
the electromagnetic field via the scale anomaly
arising from the quantum correction of the quark loop.
To obtain 
the coupling form between the scalar
field and the electromagnetic field,  
we consider 
the correlation function
related to the Ward-Takahashi identity for a photon two-point function coupled with the dilatation current $j_D^\mu$,
\begin{eqnarray}
\lim_{q\to 0}
\int d^4y e^{iq\cdot y}
\langle 0|T \partial_\rho j_D^\rho(y) A_\mu (x)A_\nu (0)|0\rangle
=
i\delta_D \langle 0| TA_\mu (x)A_\nu (0)|0\rangle
\,, 
\label{LET_EM}
\end{eqnarray}
where $\delta_D$ denotes
the infinitesimal scale transformation
by the charge $Q_D = \int d^3 {\vec x} j_D^0(x)$, defined as
$[i Q_D, {\cal O}(x)]= \delta_D {\cal O}(x)= (d_{\cal O} + x^\nu \partial_\nu) {\cal O}(x)$, for an operator ${\cal O}$ with the scaling dimension $d_{\cal O}$.

Assuming the lightest scalar-meson pole-dominance,
we can rewrite the left-hand side of Eq.~(\ref{LET_EM}),
\begin{eqnarray}
\lim_{q\to 0}
\int d^4y e^{iq\cdot y}
\langle 0|T \partial_\rho j_D^\rho(y) A_\mu (x)A_\nu (0)|0\rangle
=
-i
%
f_\phi
%
\lim_{q\to 0}
\langle \phi({q})| T A_\mu (x)A_\nu (0)| 0\rangle.
\label{lowT_DA_int}
\end{eqnarray}
Here, we have used 
Eq.~(\ref{PCDC}).
Hence,
the overlap amplitude of two photons with the chiral-singlet/isosinglet scalar state
can be described by the scale transformation for the photon propagator,
\begin{eqnarray}
\langle \phi({q=0})| T A_\mu (x)A_\nu (0)| 0\rangle
=
-
\frac{1}{f_\phi}
\delta_D \langle 0| TA_\mu (x)A_\nu (0)|0\rangle.
\label{amp_ST}
\end{eqnarray}
By using Eq.~(\ref{amp_ST}), 
later we will find
the coupling form between the sigma meson field and the electromagnetic field.
\subsection{Electromagnetic scale-anomaly induced-tadpole at finite temperature}
In the magnetized-thermal bath,
the Lorentz invariance in four dimensions is lost. Then,
the photon propagator expression should be no longer Lorentz or $O(3)$ covariant, due to the quark-loops in the bath, which would yield intrinsic 
magnetic corrections to the photon polarization function, as well as the thermal corrections.   
Note, however, that our current main concern is the overlap amplitude between 
the $\phi$ scalar and two photon fields in Eq.(\ref{amp_ST}), which, in terms of 
the interaction Lagrangian, 
should take the form like  
$\sim f(eB, T) \bar{\phi} F_{\mu\nu} F^{\mu\nu} \sim f(eB, T)\bar{\phi} B^2$, 
where $\bar{\phi}$ is the background field of $\phi$, and 
$f(eB, T)$ is an associated form factor arising from the thermomagnetized quark loop 
corrections. 
Therefore, in the weak magnetic regime that involves our 
current concern, 
we can ignore the $eB$-dependence in $f(eB, T)$, because 
it would give subleading corrections like ${\cal O}((eB)^4)$ to the total $\phi$-photon-photon amplitude. 
Hence,
as far as the weak magnetic field regime is concerned, 
the magnetic field dependence on 
the photon propagator involving polarization functions like $f(eB, T)$  
can be ignored. 
We will further give comments 
on this higher order correction later (in Conclusion and Discussion section).  

In that case, 
the photon propagator only gets the thermal quantum correction to the quark-photon vertex 
with a (dynamical) quark mass $m_{\rm dyn}$. 
Then 
the polarization structure can be generally 
decomposed into the three independent components,
\begin{eqnarray}
D_{\mu\nu}(K)
&=&
\frac{-i}{K^2-\Pi_T}P_{\mu\nu}
+\frac{-i}{K^2-\Pi_L}Q_{\mu\nu}
+(\xi-1)
\frac{iK_\mu K_\nu}{K^4},
\label{Dmunu}
\end{eqnarray}
where the four-momentum of the photon denotes
$K_\mu=(\omega,k_1,k_2,k_3)$.
$\xi$ is the gauge fixing parameter, and 
$P_{\mu\nu}$ and $Q_{\mu\nu}$ represent  
the transverse- and the longitudinal-polarization tensors,
\begin{eqnarray}
P_{\mu\nu}&=&\tilde \eta_{\mu\nu}+\frac{\tilde K_\mu \tilde K_\nu }{{\vec k}^2},
\nonumber\\
Q_{\mu\nu}&=&\frac{-1}{K^2{\vec k}^2}
\left({\vec k}^2u_\mu+\omega  \tilde K_\mu\right)
\left({\vec k}^2u_\nu+\omega  \tilde K_\nu\right),
\end{eqnarray}
with
$\tilde \eta_{\mu\nu}$, $\tilde K_\mu$ and $u_\mu$ being 
\begin{eqnarray}
\tilde \eta_{\mu\nu}&=&
{\rm diag }[0,-1,-1,-1]
,\nonumber\\
\tilde K_\mu&=& K_\mu-\omega u_\mu=(0,k_1,k_2,k_3),
\nonumber\\
u_\mu&=&=(+1,0,0,0).
\end{eqnarray}
The loop corrections $\Pi_{T,L}$ separately include
the vacuum part $\Pi_{T,L}^{(T=0)}$ and thermal part $\Pi_{T,L}^{(T\neq 0)}$,
\begin{eqnarray}
\Pi_{T,L}=\Pi_{T,L}^{(T=0)}+\Pi_{T,L}^{(T\neq0)}.
\label{PiTL}
\end{eqnarray}

At the quark one-loop calculation performed by the dimensional regularization in $4-\epsilon$ dimension, $\Pi_{T,L}$ are expressed as 
(for the detail of the thermal part, see \cite{Weldon:1982aq,Ahmed:1991mz})
\begin{align}
\Pi^{(T=0)}_T&=
\Pi^{(T=0)}_L
=
-\frac{N_c}{2\pi^2}
\sum_fQ_f^2 K^2
\int_0^1dx\, x(1-x)\left(
\frac{2}{\epsilon}-\log\left\{m^2_{\rm dyn}-x(1-x)K^2\right\}-\gamma+\log(4\pi)
\right)
 \nonumber\\
\Pi_T^{(T\neq 0)}&=
\frac{2N_c}{\pi^2}\sum_fQ_f^2
\Biggl[
\left\{
\frac{\omega^2}{{\vec k}^2}
+\left(1-\frac{\omega^2}{{\vec k}^2}
\right)\ln\frac{\omega+|{\vec  k}|}{\omega-|{\vec  k}|}
\right\}\nonumber\\
&
\times
\left(
{m_{\rm dyn}T}
\tilde a(m_{\rm dyn}/T)-T^2
\tilde c(m_{\rm dyn}/T)
\right)
+\frac{1}{8}\left\{
2m^2_{\rm dyn}+\omega^2+\frac{107\omega^2-131{\vec k}^2}{72}\right\}\tilde b (m_{\rm dyn}/T)
\Biggl]\nonumber\\
\Pi_L^{(T\neq0)}&=
\frac{4N_c}{\pi^2}\sum_f Q_f^2
\left(1-\frac{\omega^2}{{\vec k}^2}\right)
\Biggl[
\left(1-\frac{\omega}{2|{\vec k}|}
\ln\frac{\omega+|{\vec  k}|}{\omega-|{\vec  k}|}
\right)\left(
{m_{\rm dyn}T}
\tilde a(m_{\rm dyn}/T)
-T^2\tilde c(m_{\rm dyn}/T)
\right)\nonumber\\
&
+\frac{1}{4}\left\{
2m^2_{\rm dyn}-\omega^2+\frac{11{\vec k}^2+37\omega^2 }{72}\right\}\tilde b (m_{\rm dyn}/T)
\Biggl],\nonumber\\
\end{align}
where 
$Q_f$ with $f=u,d$ (and $s$) is the electromagnetic charge 
for quark flavor $f$,  $Q_u=2e/3,\;\;Q_d=-e/3,\;\;Q_s=-e/3$; $m_{\rm dyn}$ is the dynamical quark mass generated by the spontaneous chiral symmetry breaking; 
$\gamma$ is the Euler-Mascheroni constant. 
Here, we have taken the flavor universal limit; 
$m_{\rm dyn}^{(u)}=m_{\rm dyn}^{(d)}=m_{\rm dyn}^{(s)}=m_{\rm dyn}$, which turns out to be justified 
in the present framework of analysis even in the presence of the chiral-and isospin-breaking magnetic field 
(See the later discussion around Eq.(\ref{Yukawa:g})). 
$\tilde a(m_{\rm dyn}/T)$, $\tilde b(m_{\rm dyn}/T)$ and $\tilde c(m_{\rm dyn}/T)$ are given by 
\begin{eqnarray}
\tilde a(m_{\rm dyn}/T)&=&
\ln(1+e^{-m_{\rm dyn}/T}),\nonumber\\
\tilde b(m_{\rm dyn}/T)&=&
\sum_{n=1}(-1)^n {\rm Ei}(-nm_{\rm dyn}/T)
=\sum_{n=1}(-1)^n\left(
-\int^\infty_{nm_{\rm dyn}/T} \frac{e^{-t}}{t} dt
\right),
\nonumber\\
\tilde c(m_{\rm dyn}/T)&=&
\sum_{n=1}^{\infty}(-1)^n\frac{e^{-nm_{\rm dyn}/T}}{n^2}.
\end{eqnarray}

With the above explicit expression,
the scale transformation
for the photon propagator showing up in Eq.(\ref{amp_ST})
is evaluated as
\begin{eqnarray}
\delta_D D_{\mu\nu} (x)
&=&
-\int\frac{d^4p }{(2\pi)^4}\Biggl[
\Biggl\{
F^{(T=0)}(K)
+
F^{(T\neq0)}_T(\omega,{\vec k})
\Biggl\}
\frac{i}{(K^2-\Pi_T)^2}P_{\mu\nu}
\nonumber\\
&&
+
\Biggl\{
F^{(T=0)}(K)
+
F^{(T\neq0)}_L(\omega,{\vec k})
\Biggl\}
\frac{i}{(K^2-\Pi_L)^2}Q_{\mu\nu}
\Biggl]
e^{-iK\cdot x},
\label{deltaD-D}
\end{eqnarray}
where 
$F^{(T=0)}(K)$ and $F^{(T\neq0)}_{T,L}(\omega,{\vec k})$ represent 
the variation for
the vacuum- and the thermal-correction parts, respectively. 
The vacuum part $F^{(T=0)}(K)$ is evaluated as 
\begin{eqnarray}
F^{(T=0)}(K)&=&-2\frac{\beta(e)}{e}
K^2
+
F_{\rm NL}^{(T=0)}(K)
, 
\label{func_vacuum_beta}
\end{eqnarray}
where 
\begin{eqnarray}
F_{\rm NL}^{(T=0)}(K)=\frac{N_c}{\pi^2}
\sum_f Q_f^2
m^2_{\rm dyn}
\Biggl[-1
+
\int_0^1 dx\frac{1}{1-x(1-x)(K^2/m^2_{\rm dyn})}\Biggl].
\end{eqnarray}
$\beta(e)$ denotes the beta function of the electromagnetic gauge coupling $(e)$, defined 
as $\beta(e)= \partial e(\mu)/\partial \ln \mu $ with the renormalization scale $\mu$.  
We evaluate 
the beta function $\beta(e)$ 
at the one-loop level of QED as    
$ 
\beta(e)=\frac{e}{(4\pi)^2}\frac{4N_c}{3}
\sum_f Q_f^2 
$, with $N_c=3$. 
The first term of the right hand side in Eq.~(\ref{func_vacuum_beta}) 
induces the local-tadpole interaction between the $\phi$  and the electromagnetic fields, 
while the second term $F_{\rm NL}^{(T=0)}(K)$ corresponds to the nonlocal effective interaction.



The thermal-correction part $F^{(T\neq 0)}_{T,L}(\omega,{\vec k})$ in Eq.(\ref{deltaD-D}) gives the 
the effective interaction between the $\phi$ and the electromagnetic field, which is expressed as the  nonlocal form.  
Note, however, that $F^{(T\neq 0)}_{L}(\omega,{\vec k})$ comes along with 
the longitudinal polarization of photon, which  
does not couple to a magnetic field, so that it does not contribute to the the thermomagnetic tadpole.  Hence this part will be discarded in the present study.

From Eq.~(\ref{amp_ST}) and Eq.~(\ref{deltaD-D}),
we can read off the effective interaction between the $\phi$ and the electromagnetic field at finite temperature: 
\begin{eqnarray}
\int d^4 x
{\cal L}_{\rm int}^{\rm (Tad)}&=&
\int d^4 x
\frac{\beta (e)}{2e}
\frac{\phi}{f_\phi}
F_{\mu\nu} F^{\mu\nu}
\nonumber\\
&&
+\int d^4 x \int d^4 y
f^{(T=0)}(x-y) \frac{\phi(x)}{f_\phi}
{F}_{\mu\nu}(x) {F}^{\mu\nu}(y)
\nonumber\\
&&
+\int d^4 x \int d^4 y
 f^{(T\neq0)}(x-y)
\frac{\phi(x)}{f_\phi}
\bar{F}_{\mu\nu}(x) \bar{F}^{\mu\nu}(y) 
,
\label{EM_SA_int_v2}
\end{eqnarray}
where
\begin{eqnarray}
\bar F_{\sigma \rho}(x)&=&\tilde\partial_\sigma  \tilde A_\rho(x)
-
\tilde \partial_\rho  \tilde A_\sigma(x),
\nonumber\\
\tilde \partial_\sigma 
&=&\partial_\sigma - u_\sigma \partial_0
=(0, \partial_1,\partial_2,\partial_3),
\nonumber\\
\tilde A_\rho(x)&=&\tilde \eta_{\rho \mu } A^\mu(x) 
=A_\rho (x)-A_0(x) u_\rho
=(0,A_1,A_2,A_3)\,, \notag\\ 
\,
f^{(T=0)}(x-y)&=&
\int \frac{d^4K}{(2\pi)^4}
\frac{-1}{4 K^2}
F^{(T=0)}_{NL}(K)
e^{-iK\cdot (x-y)},
\nonumber\\
f^{(T\neq0)}(x-y)&=&
\int \frac{d^4K}{(2\pi)^4}
\frac{1}{4\vec k^2}
F^{(T\neq0)}_T(\omega,\vec k)
e^{-iK\cdot (x-y)}.
\end{eqnarray}

We readily see that at $T=0$, the first term in Eq.~(\ref{EM_SA_int_v2}) corresponds to the electromagnetic scale anomaly in the vacuum, which induces the tadpole term 
in the effective potential viewed as a Ginzburg-Landau description.
The second term of the nonlocal interaction vanishes at the low energy limit, so that   
it does not contribute to the effective potential describing the chiral phase transition.
On the other hand, the third term corresponding to 
the thermal-correction part in Eq.(\ref{EM_SA_int_v2}) does not drop out even in the low-energy limit, 
and includes an infrared  
divergence, which we regularize by introducing the cutoff $\mu_{\rm IR}$.   
Then, in the constant magnetic field $eB$, the thermomagnetically induced tadpole can be evaluated as 
\begin{align}
\int d^4 x \int d^4 y
 f^{(T\neq0)}(x-y)
\frac{\bar\phi}{f_\phi}
\bar{F}_{\mu\nu}(x) \bar{F}^{\mu\nu}(y) 
&=
\int d^4 x \int d^4 y
 f^{(T\neq0)}(x-y)
\frac{\bar\phi}{f_\phi}
2B^2
\nonumber\\
&\approx
\int d^4x
\left[
\frac{1}{2e^2}
\sum_fQ_f^2
F(T,m_{\rm dyn})
\frac{\bar\phi}{f_\phi}
\frac{|eB|^2}{\mu_{\rm IR}^2}
\right],
\label{thermal-tad-part}
\end{align}
where
\begin{align}
\sum_f Q_f^2F(T,m_{\rm dyn})
&=F^{(T\neq0)}_T(\omega=0,{\vec k}=0)\nonumber\\
&=
\sum_f Q_f^2
\frac{4N_c}{\pi^2}
\Biggl[
\left(
{m_{\rm dyn}T}
\tilde a(m_{\rm dyn}/T)
-T^2\tilde c(m_{\rm dyb}/T)
\right)
+\frac{1}{4}m^2_{\rm dyn}\tilde b (m_{\rm dyn}/T)
\Biggl].
\end{align}

In a weak magnetic field regime 
(with $\sqrt{eB} \ll  {\cal O}$(GeV), much less than an ultraviolet cutoff scale for an effective model which we will work on in the later 
sections)  
the magnetic field strength $\tilde F_{12}=-B$ should 
numerically supply the infrared cutoff, i.e., $\mu_{\rm IR} \equiv \sqrt{eB}$. 
With this prescription applied, 
the right hand side of Eq.(\ref{thermal-tad-part}) takes the form
\begin{eqnarray}
\int d^4x
\left[
\frac{1}{2e^2}
\sum_fQ_f^2
F(T,m_{\rm dyn})
%
\frac{\bar\phi}{f_\phi}
|eB|
\right].
\end{eqnarray}
In total,
the tadpole term is generated in the effective potential for the background field $\bar \phi$ 
and 
takes the form 
\begin{eqnarray}
{V}_{\rm eff}^{\rm (Tad)}(\bar \phi)
=
-
%
\frac{\bar \phi}{f_\phi}
\Biggl[
\frac{\beta (e)}{e}
B^2
+\frac{1}{2e^2}
\sum_fQ_f^2
F(T,m_{\rm dyn})
|eB|
\Biggl].
\label{EM_tad_appendix}
\end{eqnarray}

As the magnetic field gets strong, the one-loop calculation depending the magnetic field should be taken into account. However, the
dynamics of quarks are governed by the lowest Landau
level states polarized along the direction parallel to the
magnetic field, where
the transverse part of the photon polarization does not couple to quarks. Thus, the tadpole term induced by the electromagnetic scale anomaly Eq.~(\ref{EM_tad_appendix}) vanishes in the strong magnetic field regime, which is most effective in the weak magnetic regime.


In Eq.~(\ref{EM_tad_appendix}),
the $\phi$ can be identified as the  
the sigma meson which plays a role of the order parameter for the spontaneous chiral symmetry breaking.   
In terms of the linear sigma field $\Phi$, 
we find $\phi = \sqrt{2 {\rm tr} [\Phi^\dag \Phi]} = \sqrt{ \sigma_0^2 + \pi^2 + \cdots}$, and $\bar{\phi}= \bar{\sigma}_0$ 
when we only focus on the $\sigma_0$-direction, which is relevant to 
the chiral phase transition. 
Then the $\phi$-tadpole term in
Eq.~(\ref{EM_tad_appendix}) can be expressed as
\begin{eqnarray}
{V}_{\rm eff}^{\rm (Tad)}(\bar \sigma_0)
=
-
%
\sqrt{\frac{2}{N_f}}
\frac{\bar \sigma_0}{f_\pi}
\Biggl[
\frac{\beta (e)}{e}
B^2
+\frac{1}{2e^2}
\sum_fQ_f^2
F(T,m_{\rm dyn})
|eB|
\Biggl], 
\label{EM_tad_PhiBase}
\end{eqnarray} 
where 
we have used $ f_\phi = \sqrt{N_f/2} f_\pi$.  
Hence, the scale-anomaly induced-tadpole terms in Eq.~(\ref{EM_tad_PhiBase}) give the contributions to the chiral phase transition. 
This thermomagnetic tadpole arises even in the chiral limit, as 
the magnetically induced-explicit chiral-breaking effect.

Before closing this section, we give a comment on 
possible contribution from the gluonic scale-anomaly, 
which is thought to be present all the way in the thermal QCD including the 
vacuum. 
It is known that the gluonic scale-anomaly can be introduced as the log potential form in the scale-chiral Lagrangian~\cite{Brown:1991kk}, which satisfies the low energy theorem in Eq.~(\ref{PCDC}). 
In the linear sigma model, 
however, the meson terms already saturate the low energy theorem, i.e., 
the scale anomaly and the partially-conserved dilatation current relation: 
\begin{align}
\partial_\mu j^\mu_D=
\left(T^\mu_{\;\;\mu}\right)_{\rm meson}
&=
2\mu^2{\rm tr}[\Phi^\dagger \Phi]
+(N_f-4)B({\rm det \Phi}+{\rm det \Phi^\dagger})
-3c \, {\rm tr}[{\cal M}\Phi^\dagger+{\cal M}^\dagger \Phi],
\nonumber\\ 
\to \qquad 
\langle 0|
\left(T^\mu_{\;\;\mu}\right)_{\rm meson}(x)|{\sigma}_0(p) \rangle 
&=
\sqrt{\frac{N_f}{2}}
f_\pi m^2_\sigma  
e^{-ip\cdot x}
. 
\label{PCDC_linear}
\end{align} 
Thereby, such a conventional log interaction term will be redundant. 
However, 
one might still think how much the log potential form like $\bar \sigma_0^4 \log\bar \sigma_0 $ is responsible for the chiral phase transition.
Actually, the log potential can be rewritten as a combination of $\bar \sigma_0^{4}$ term and $\bar \sigma_0^{4+\epsilon}$ term with a small $\epsilon,$  $\epsilon\ll1$.
Therefore,
the log potential does not give any interference for 
the electromagnetically induced-scale anomaly tadpole term. 
Thus the tadpole term cannot be washed out and is intact even if we include the redundant log potential into the effective potential.

\section{A new critical endpoint: demonstration based on quark meson model in the large $N_c$ limit}

To explicitly see what form the tadpole takes, 
we employ a quark-meson model in a constant weak magnetic field~\footnote{
It has been discussed~\cite{Ayala:2014gwa,Andersen:2014oaa} that 
in a strong magnetic field regime, the quark-meson model  
can reproduce the reduction of 
the (pseudo) critical temperature of the chiral crossover 
and the inverse magnetic catalysis for the quark condensate,  
discovered on the lattice~\cite{ catalysis2, endrodi-eos, Tomiya-Hisq, Massimo-heavy-pion, Enrodi-heavy-pion,Endrodi}. 
}. 
The model Lagrangian is built based upon 
renormalizable interactions among quarks and mesons 
allowed by the chiral invariance. 
It is constructed from the linear sigma model part in Eq.(\ref{Lag:LSM}) 
and the Dirac-fermion kinetic term for quarks, together with  
the Yukawa interaction term between quarks and mesons. 
The Yukawa coupling is introduced to be flavor 
universal, as it should be, because QCD is flavor blind: 
\begin{align} 
 {\cal L}_{\rm Yukawa}= - \sum_q g_q (\bar{q}_L \Phi q_R + {\rm h.c.}) 
\qquad {\rm with} \qquad g_q \equiv g  
\,. \label{Yukawa:g}
\end{align} 
Thus, the quark-meson model Lagrangian goes like 
\begin{align} 
{\cal L}_{\rm QM} = {\cal L}_{\rm LSM} + {\cal L}_{\rm kin}^{\rm quarks} + {\cal L}_{\rm Yukawa} 
\,. 
\end{align}
As in the literature~\cite{Andersen:2011ip}, we may work in 
the large $N_c$ limit, and only take into account the quark loops 
at one-loop level.

As in Eq.(\ref{GL-model}), we take  
the flavor symmetric order parameter,  
${\bar\sigma}_u = \bar{\sigma}_d = \bar{\sigma}_s = \bar{\sigma}_0$, though potentially non-negligible flavor 
breaking can be induced by the magnetic field. 
This prescription can be justified as follows: 
as it will be seen, the later concrete analysis 
will be based on the quark-meson model 
at the large $N_c$ limit, where among the 
charged particle contributions, 
only the quark loops via the flavor-universal Yukawa coupling in Eq.(\ref{Yukawa:g})
contribute 
to the chiral order parameter at vacuum ($T=eB=0$). 
Thus, in the thermomagnetic system with massless three or two flavors, 
only the charge difference among three or two quarks 
makes the chiral order parameters flavorful, 
which comes with the magnetic field. 
Meanwhile, lattice QCD (with 2 + 1 flavors) at physical point has reported small isospin breaking in the up and 
down quark condensates at around the chiral crossover, 
even in a strong magnetic field ($\sqrt{eB} \ge 1$ GeV), 
e.g., ~\cite{Endrodi:2015oba}. 
This small flavor breaking can be applied also 
among up and strange quarks with a weaker magnetic field in the present analysis. 
Thus we simply assume the three-flavor symmetric vacuum 
even in the presence of magnetic field.

In the large $N_c$ limit,
the quantum correction to the effective potential 
only arises from the 
quark one-loop calculation
regularized by the dimensional regularization, which consists of
the vacuum part $V_{\rm 1-loop}^{\rm vac}$
and the thermal part
$V_{\rm 1-loop}^{\rm T}$~\cite{Andersen:2011ip}.
By combining the quark parts with 
the mesonic part in Eq~(\ref{tree_pot}) and 
the scale-anomaly
induced-tadpole term in Eq.~(\ref{EM_tad_PhiBase}),
the effective potential based on the quark-meson model
is given as,
\begin{eqnarray}
V_{\rm eff}(\bar \sigma_0)
=V_{\rm tree}(\bar \sigma_0)+
V_{\rm 1-loop}^{\rm vac}(\bar \sigma_0)+
V_{\rm 1-loop}^{\rm T}(\bar \sigma_0)+
{V}_{\rm eff}^{\rm (Tad)}(\bar \sigma_0),
\end{eqnarray}
where
\begin{eqnarray}
V_{\rm 1-loop}^{\rm vac}&=&
\frac{N_c m_{\rm dyn}^4}{(4\pi)^2}\sum_f\left[
\log\frac{\Lambda^2}{2|Q_fB|}+1
\right]
-\frac{N_c}{2\pi^2}\sum_f (Q_f B)^2  \left[
\zeta^{(1,0)}(-1,x_f)+\frac{1}{2}x_f\log x_f
\right]\nonumber\\
V_{\rm 1-loop}^{T}
&=&
-N_c\sum_{s,f,n}\frac{|Q_fB|T}{\pi^2}
\int^\infty_0 dp\log\left(
1+e^{-\sqrt{p^2+
\left(M_{\rm eff}^{(f,s)}\right)^2
}/T   }
\right)
\nonumber\\
&=&
-N_c\sum_f\frac{|Q_fB|T}{\pi^2}
\int^\infty_0 dp
\Biggl[
\log\Biggl\{
1
+\exp\left ({-\sqrt{p^2+\left(m^{(f)}_{\rm dyn}\right)^2}}/T\right)
\Biggl\}
\nonumber\\
&&
+2\sum_{n=1}
\log\Biggl\{
1
+\exp\left ({-\sqrt{p^2+\left(m^{(f)}_{\rm dyn}\right)^2+2n|Q_f B|}}/T\right)
\Biggl\}
\Biggl],
\end{eqnarray}
with
\begin{eqnarray}
M_{\rm eff}^{(f,s)}
&=&
\sqrt{\left(m^{(f)}_{\rm dyn}\right)^2+|Q_fB|(2 n+1-s)}
\nonumber\\
x_f&=&{\left(m^{(f)}_{\rm dyn}\right)^2}/{(2|q_f B|)}.
\end{eqnarray}
$M_{\rm eff}^{(f,s)}$ represent the effective quark masses 
for the spin-up/down quark, $s=\pm1$.
$\zeta(a,x)$ is the Hurwitz zeta-function, and
$\Lambda$ being the renormalization scale
fixed as $\Lambda=181.96{\rm MeV}$~\cite{Andersen:2011ip}.
As noted above, 
the dynamical quark masses as well as the quark condensates 
are flavor universal, and are evaluated as 
\begin{eqnarray}
m^{(u)}_{\rm dyn}
=
m^{(d)}_{\rm dyn}
=
m^{(s)}_{\rm dyn}
=
m_{\rm dyn}=
2g\bar \Phi, 
\end{eqnarray}
where
 $g$ is the Yukawa coupling between the quarks and the mesonic sector as in Eq.(\ref{Yukawa:g}).
 At $T=0$ and $eB=0$, 
 we have set $m_{\rm dyn}=300\;{\rm MeV}$ and $g=3.2258$~\cite{Andersen:2011ip}, 
 which produces meson mass values given in Appendix~A.


\begin{figure}[ht]
\begin{tabular}{cc}
 \begin{minipage}{0.5\hsize}
\begin{center}
   \includegraphics[width=7.0cm]{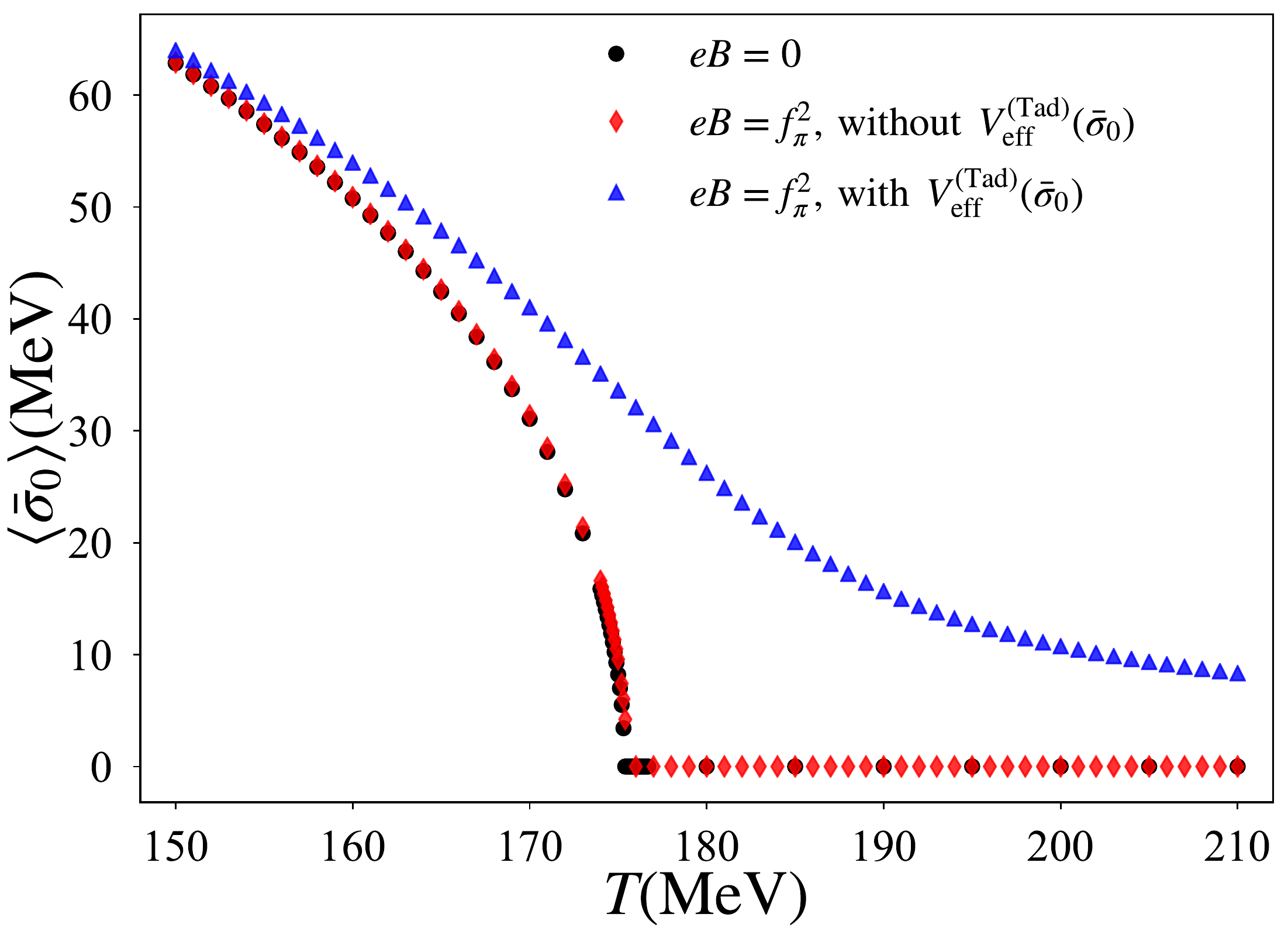}
    \subfigure{(a)}
  \end{center}
 \end{minipage}%
 \begin{minipage}{0.5\hsize}
  \begin{center}
   \includegraphics[width=7.0cm]{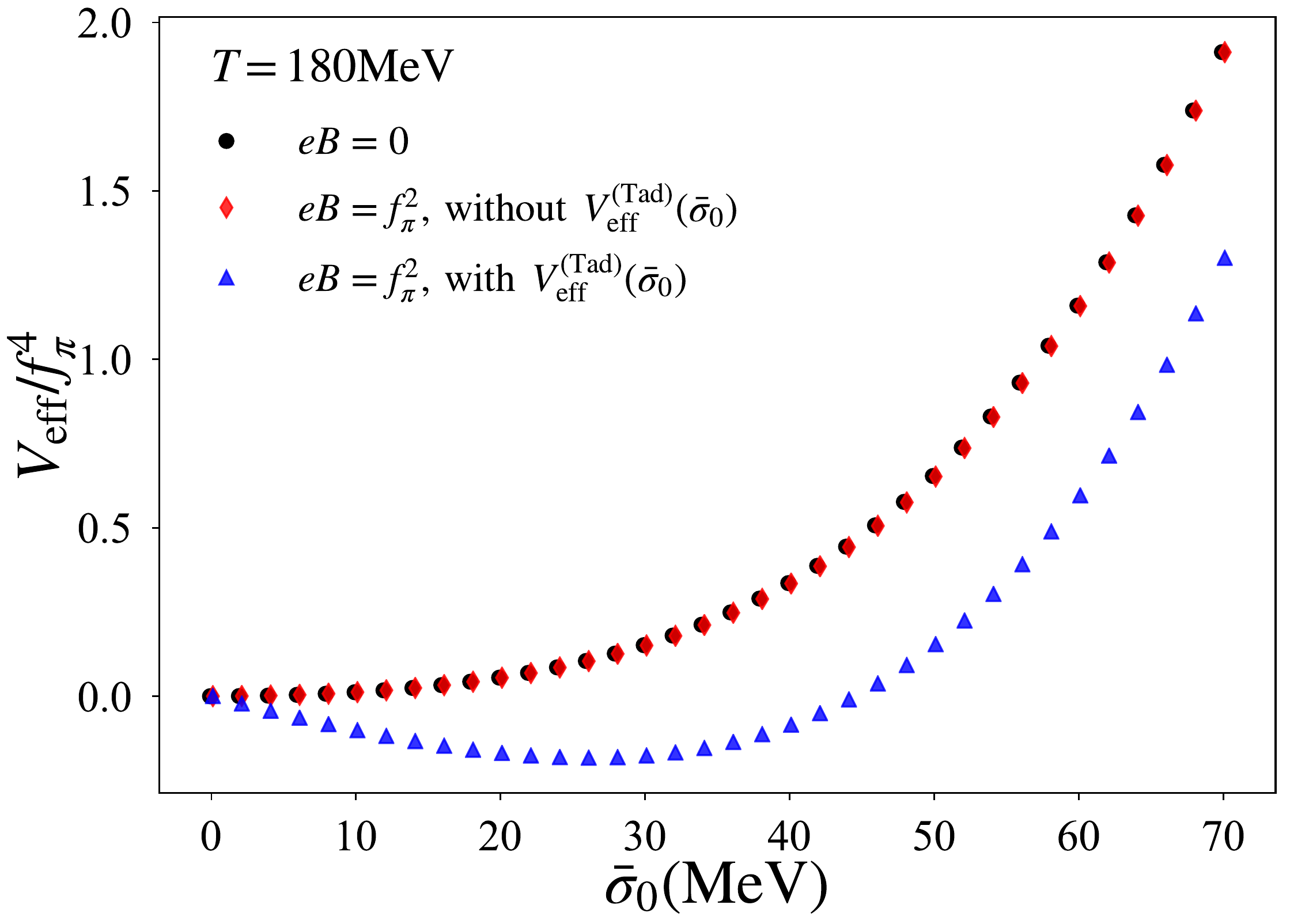}
    \subfigure{(b)}
  \end{center}
 \end{minipage}
  \end{tabular}
 \caption{
 The chiral phase transition for massless two-flavor case, predicted from the quark-meson model in the large $N_c$ limit. 
 The panel (a) shows the $T$-dependence of the chiral order parameter $\bar{\sigma}_0$, around the critical 
 temperature $T_c \simeq 175$ MeV. 
 The panel (b) displays the deformation of the 
 effective potential $V_{\rm eff}$, 
 normalized to the (forth power of) pion decay constant $f_\pi (\simeq 93 \,{\rm MeV})$ at vacuum. 
 The effective potential has been normalized by subtracting  
 $V_{\rm eff}(\bar{\sigma}_0=0)$ 
 at $T=180$~MeV. 
 Both panels monitor the $eB$ dependence in unit of 
 $f_\pi^2$, for two cases with or without the scale 
 anomaly-induced tadpole term included.  
 }  
 \label{nf2}  
\end{figure}


Two plots in Fig.~\ref{nf2} 
exhibit a sketch of the thermomagnetic phase transition for the chiral $SU(2)_L \times SU(2)_R$ symmetry based on the quark-meson model with massless two flavors.  
As clearly seen, the scale-anomaly induced-tadpole term in Eq.(\ref{EM_tad_PhiBase}) makes the phase transition 
crossover, in somewhat a weak magnetic regime 
less than the (square of) chiral critical temperature $T_c \simeq 175$ MeV (for $eB=0$)~\footnote{
 The value of $T_c$ estimated from the present quark-meson model does not agree with the result with the chiral extrapolation (only applied to the lightest two-flavors)  
 on the lattice QCD with $2 + 1$ flavors, 
 $T_c|_{\rm lat.} = 132^{+3}_{-6} $MeV \cite{Sheng-Tai}. 
 However, this quantitative discrepancy is 
 irrespective of our main claim in Fig.~\ref{fig:columbia_plot}.
\label{foot:Tpc} }. 
This puts the milestone marked as ``Crossover" 
at the crossed point in the $eB$ axis 
at $m_s \to \infty$, in the extended Columbia plot 
in Fig.~\ref{fig:columbia_plot}.


\begin{figure}[ht]
\begin{tabular}{cc}
 \begin{minipage}{0.5\hsize}
\begin{center}
   \includegraphics[width=7.0cm]{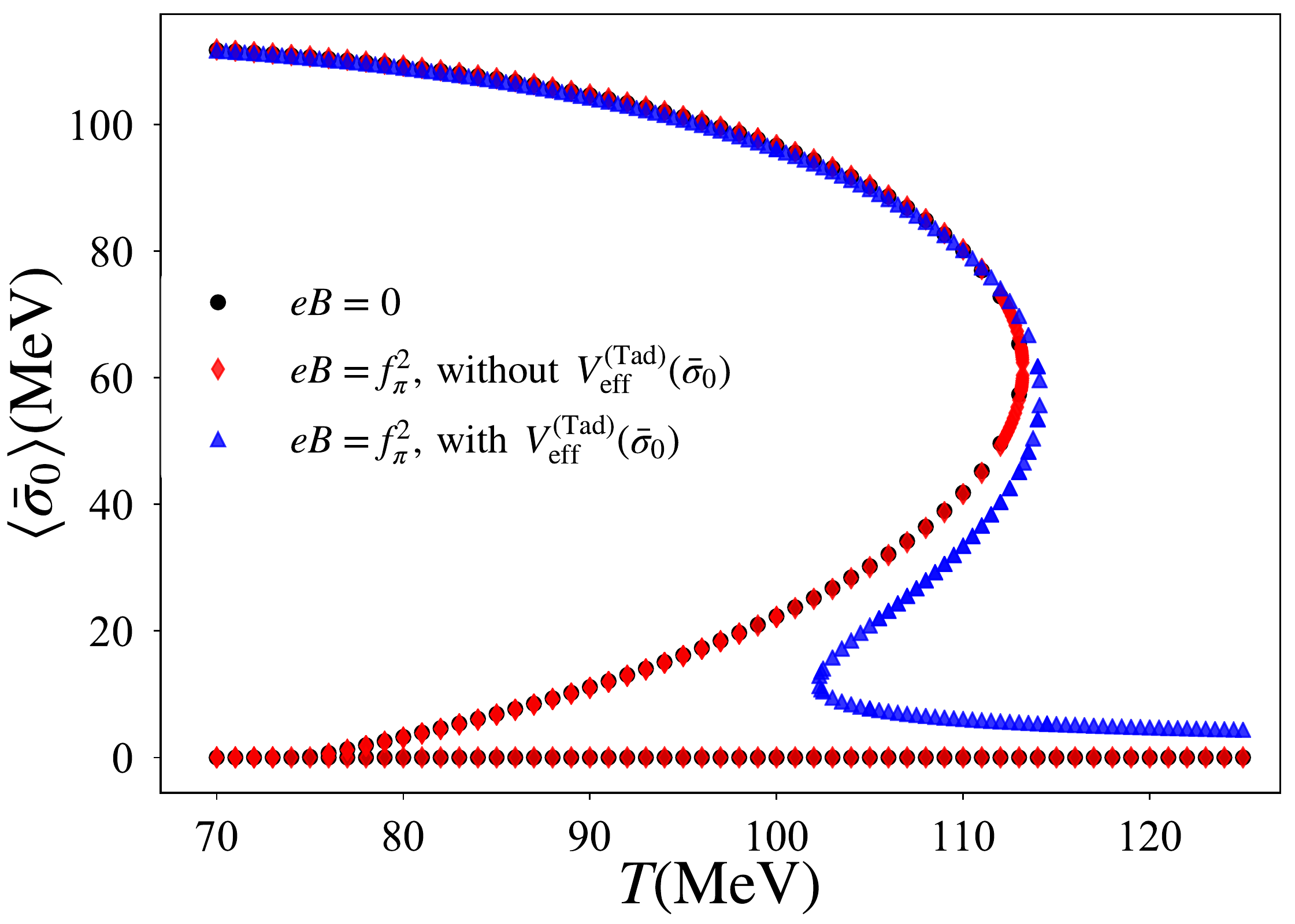}
    \subfigure{(a)}
\end{center}
 \end{minipage}%
 \begin{minipage}{0.5\hsize}
 \begin{center}
   \includegraphics[width=7.0cm]{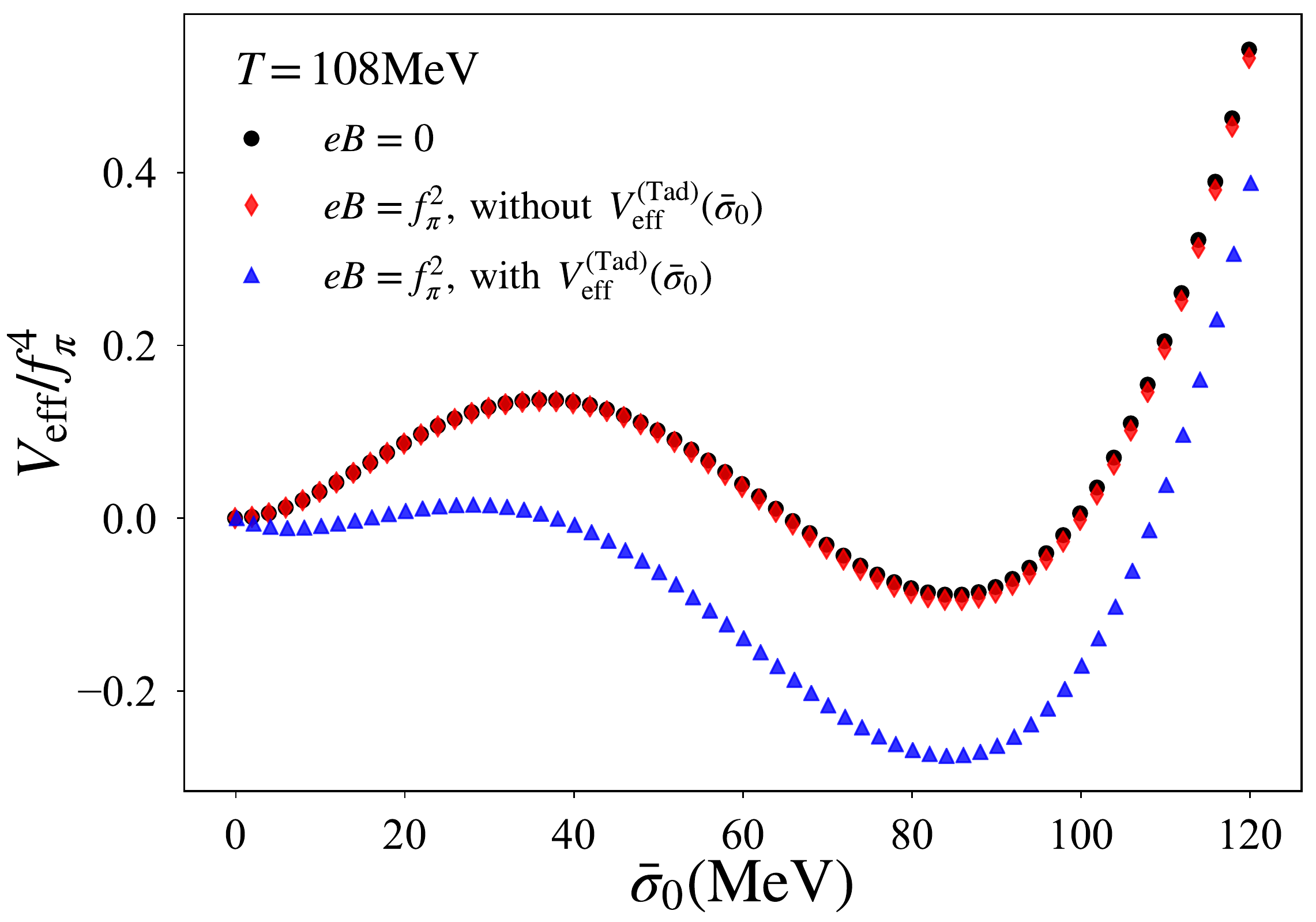}
    \subfigure{(b)}
  \end{center}
   \end{minipage}
\end{tabular}
 \caption{
 The same as Fig.~\ref{nf2}, but for massless three flavors. The chiral critical temperature 
 at $eB=0$ is 
 $T_c=113$~MeV. The effective potential $V_{\rm eff}$ has been normalized in the same way as 
 in Fig.~\ref{nf2}, and subtracted by the value at $\bar{\sigma}_0=0$ evaluated at $T=108$ MeV. 
 }  
 \label{nf3}  
\end{figure}


Turning to the massless three-flavor case, 
we make similar plots in Fig.~\ref{nf3}. 
In this case, the electromagnetically induced scale-anomaly tadpole 
competes with the $U(1)$ axial-anomaly induced 
cubic term.  
Eventually the $U(1)$ axial-anomaly contribution wins, so that 
the chiral phase transition 
for the $SU(3)_L \times SU(3)_R$ symmetry 
keeps still of the first order, 
even the presence of the induced tadpole,
 though the chiral order parameter 
 cannot exactly reach zero because of the tadpole. 
This result is marked as ``1st order" in the 
extended Columbia plot, in Fig.~\ref{fig:columbia_plot}. 
It has been noted in the literature~\cite{Pisarski:2019upw} 
that at $T < 1.5 T_c$ the instanton effects are not 
significantly suppressed. In terms of the present quark-meson 
model, this implies that the corresponding coupling $B_{\rm anom}$ in Eq.~(\ref{Vanom}) for the $U(1)$-axial anomaly term,  
giving the $\alpha_3(T,eB) \bar{\sigma}_0^3$ term in Eq.(\ref{GL-model}), 
can still be effective over the 
possible thermal suppression at around $T=T_c$.

Thus, combining the explicit results in Figs~\ref{nf2} and~\ref{nf3}, 
we can conclude that 
in the extended Columbia plot, 
along the 
$m_s$ axis, 
there should be a critical endpoint marked as 
a second order phase transition -- that is 
the ``Suggested endpoint", in Fig.~\ref{fig:columbia_plot} 
dictating the critical contribution of 
the electromagnetic scale anomaly to the chiral 
phase transition.

\section{Conclusion and Discussion}

In conclusion, 
detection of the scale anomaly in the chiral phase transition 
is possible: 
apply a weak magnetic field onto QCD with $m_s \to \infty$ 
and $m_u=m_d=0$, first observe the chiral crossover. 
Second, adjusting  
$m_s \searrow 0$,  
observe the second-order chiral phase transition, 
which is a new critical endpoint 
(See Fig.~\ref{fig:columbia_plot}). 
That is the evidence of the quantum scale anomaly.  
This is most operative in a weak magnetic field regime.

Several comments are in order: 

\begin{itemize}

\item 

Looking at Fig.~\ref{fig:columbia_plot}, we may have  
one interesting question:  
whether or not the second order nature persists when 
$m_s$ gets smaller along the $m_s$ axis.   
If it does, there should be a second-order phase transition 
line on the $m_s$ axis, extended from the suggested 
endpoint down to the first-order phase transition domain 
surrounded by the observed first-order point in Fig.~\ref{nf3}. 
This is illustrated as the black-bold line on the $m_s$ axis, 
in Fig.~\ref{fig:columbia_plot}. 
Here one should notice that there should also be 
a tri-critical point, at which the three phases (first, second, and crossover 
domains) overlap. It is marked by a blank square, in Fig.~\ref{fig:columbia_plot}. 
In that case, existence of the second-order phase transition line might be 
interpreted a remnant of the universality class of $O(4)$ on the $m_s$ axis, 
as in the Columbia plot without magnetic field~\cite{Columbia}.

Another second order line extended to the regime with $m_{l}>0$ 
is assumed to survive as in the usual Columbia plot without magnetic field, 
though it is yet unclear to be governed by the universality class of $Z_2$  
(See the black-bold line separating the first order and crossover domains around the massless 
three-flavor regime in Fig.~\ref{fig:columbia_plot}). 
If the crossover nature, confirmed at $m_s \to \infty$ in Fig.~\ref{nf2},  
keeps until the first-order phase transition domain 
surrounded by the observed first-order point in Fig.~\ref{nf3}, 
the new critical endpoint would then become identical with the tri-critical point. 
Detailed analysis on identification of the tri-critical endpoint is to be 
pursued in another publication.

\item As to the systematic expansion with respect to the magnetic 
field, we have applied the standard way as has been applied on lattice QCD, e.g.~\cite{DElia:2011koc,Tomiya-Hisq}. 
When expanded formally in amplitudes including 
the vacuum polarization function of photon,  
the magnetic field is assumed to be small compared to 
a cutoff scale intrinsic to the model, which is the lattice 
scaling in the case of lattice QCD, and for the present 
quark-meson model it is an ultraviolet scale of ${\cal O}(1)$ GeV, above which mesons should dominantly be melted, or gluonic degrees of freedom should become relevant. 

\textcolor{black}{
In Eq.(\ref{EM_tad_PhiBase}) we have presently ignored the next-to-leading order terms of  ${\cal O}(eB^2)$ for the thermally induced tadpole part. 
However, such higher order corrections would naively 
be suppressed by a loop factor, compared to the leading order term of 
${\cal O}(eB)$ in Eq.(\ref{EM_tad_PhiBase}). 
Even if it would not sufficiently be suppressed, and 
would potentially constructively enhance or destructively weaken the tadpole contribution, 
the presently addressed chiral-phase transition nature 
would not substantially be altered, as long as the magnetic field 
is weak enough, 
hence existence of the new critical endpoint would still be stable. }

To be more rigorous, it is subject to nonperturbative analysis on  
the photon polarization function coupled to the sigma field. 
Recently,  
there have been attempts to compute the photon polarization
in the weak field approximation
at finite temperature
\cite{Ghosh:2019kmf}. 
It would be interesting to apply their technique described 
in the literature to our study, which is, however, beyond the scope of the present paper, instead deserve to another work.

\item 
Currently, the chiral-extrapolation on lattice 
in absence of a magnetic field 
has extensively been systematically studied, and improved so much  
in the case with $2+1$ flavors \cite{Sheng-Tai}. 
On the other hand, 
there has been remarkable progress on creating small magnetic 
fields at the physical point, and the current lower bound is 
$\sim$ 100 MeV \cite{Endrodi}. 
It has also been worked out 
to calculate derivatives with respect to magnetic fields 
directly on the lattice, so in principle small magnetic fields become accessible~\cite{Bali:2020bcn}. 
Therefore, it is highly anticipated that 
the new critical endpoint is accessible to search 
on lattice QCD in near future.

\item 
More on thermodynamical properties in thermomagnetic QCD 
with the electromagnetic scale anomaly 
will be pursued in another publication. 
It is also worth cross-checking our finding 
by the (nonperturbative) renormalization group 
method.

\item The presently proposed new critical endpoint is off the physical point. 
Extension by including a finite baryon chemical potential ($\mu_B$)
might kick it up at the physical point, analogously to the 
prospected existence of the QCD critical endpoint at the physical point in 
the conventional QCD phase diagram on the $\mu_B$-$T$ plane: the QCD critical endpoint at the physical point can be  extended to form a critical surface when three quark masses 
are varied. 
This critical surface may also cover (or sweep) the smaller mass regime in evolving $\mu_B$ down to $\mu_B=0$, 
onto the Columbia plot, which is governed by the first-order domain stemming from  
the massless three-flavor limit~\cite{deForcrand:2017cgb} (See also  the ``1st order" domain in  Fig.~\ref{fig:columbia_plot}). 
Thus 
the existence of the QCD critical endpoint can be understood 
by a simple extrapolation along the critical surface with finite $\mu_B$ , and can be interpreted as weakening of the first-order nature of 
the three-flavor chiral limit to the physical point 
on the Columbia plot~\cite{deForcrand:2017cgb}. 
This kind of critical surface might also be observed 
so as to link the suggested off-physical endpoint in Fig.~\ref{fig:columbia_plot}, to the physical point, 
when 
extended to finite $\mu_B$. 
Investigation along this possibility would also be worth pursuing.

\item 

Beyond the chiral limit, 
we have also studied the pion-mass ($m_\pi$)
dependence on the chiral-phase transition nature, 
in light of lattice simulations in the future. 
Readers having particular concern in this research direction 
can refer to the result displayed in Fig.~\ref{nf3:mpi-changed}, in Appendix B.


\item Our finding would also be relevant to modeling magnetized thermal QCD-like theories beyond the 
standard model of particle physics. 
For instance, it would impact gravitational wave productions 
addressed by a chiral phase transition in a dark/hidden QCD theory 
with three flavors together with a magnetic field weaker than 
the target QCD scale, which originates from redshifting 
the primordially produced one. 
Having one heavier flavor among three would be incompatible and 
cannot realize the first-order phase transition desired to 
create sources of gravitational waves.

\end{itemize}

At any rate, searching for the new critical endpoint is 
of importance, and paves a way to pioneer this frontier 
along with the scale anomaly, 
in the 
extended Columbia plot, Fig.~\ref{fig:columbia_plot}.

\section*{Acknowledgements}

\vspace{15pt}
We are grateful to Massimo D’Elia, Heng-Tong Ding and Gergely Endr\"odi for useful comments. 
This work was supported in part by the National Science Foundation of China (NSFC) under Grant No.11747308, 11975108, 12047569, 
and the Seeds Funding of Jilin University (S.M.).  
The work of A.T. was supported by the RIKEN Special Postdoctoral Researcher program
and partially by JSPS  KAKENHI Grant Number JP20K14479.
\appendix

\section{Meson masses in linear sigma model}

In this Appendix, we list the mass formulae of linear-sigma model mesons at tree level 
(based on the model Lagrangian Eq.(\ref{Lag:LSM})), 
which are used to fix the model parameters in the main text.  
The mass formulae at the tree-level will still be available 
in the large $N_c$ limit, where only the quark loop effects 
are taken into account.

For the two-flavor case,
the meson masses are: 
\begin{eqnarray}
m_\pi^2&=&\mu^2
+\frac{1}{2}\lambda_1f_\pi^2
\nonumber\\
&=&
\sum_f c m_f
\frac{1}{f_\pi},
\nonumber\\
m_\sigma^2
&=&\mu^2+\frac{3}{2}\lambda_1f_\pi^2.
\end{eqnarray}
Here, we have used the stationary condition in Eq.(\ref{sc}).
In addition, 
$B$($\lambda_2$) has been absorbed into $\mu^2$($\lambda_1$), 
as noted above. 
In the chiral limit ($m_f \to 0$), the pion mass $m_\pi$ goes to zero.   
To implement the numerical calculation 
in the present paper for the two-flavor chiral limit, 
we have 
set  
$m_\sigma=800$ MeV, just for a reference value 
as in the  literature~\cite{Andersen:2011ip,Andersen:2014oaa}. 

For the three-flavor case, 
meson masses are: 
\begin{eqnarray}
m_\pi^2&=&
\sum_f c m_f
\sqrt{\frac{2}{3}}\frac{1}{f_\pi}
\nonumber\\
m_\sigma^2&=&
\mu^2
-B_{\rm anom}
f_\pi
+
\frac{3}{2}
(\lambda_1+3\lambda_2)
f_\pi^2
.
\nonumber\\
m_{\eta'}^2
&=&m_\pi^2+
\frac{3}{2}
B_{\rm anom} 
f_\pi
\nonumber\\
m_{a_0}^2&=&
\frac{2}{3}m_{\eta'}^2
+\frac{1}{3}
m_\pi^2
+
\lambda_1
f_\pi^2
\nonumber\\
&=&
m_\sigma^2+m_{\eta'}^2-m_\pi^2
-
3
\lambda_2
f_\pi^2
.
\end{eqnarray}
For the three-flavor chiral limit analysis,   
meson masses have been fixed to 
$m_\sigma=650$ MeV,
$m_{a_0}=940$ MeV,
and $m_{\eta'}=960$ MeV, which are typical values obtained 
in the framework of linear sigma  models~\cite{Ishida:1999qk,Kuroda:2019jzm}. 

\section{
Chiral phase transition in 
three flavor symmetric limit} 

\label{three-flavor-limit}

\begin{figure}[t]
\begin{tabular}{cc}
 \begin{minipage}{0.5\hsize}
\begin{center}
   \includegraphics[width=7.0cm]{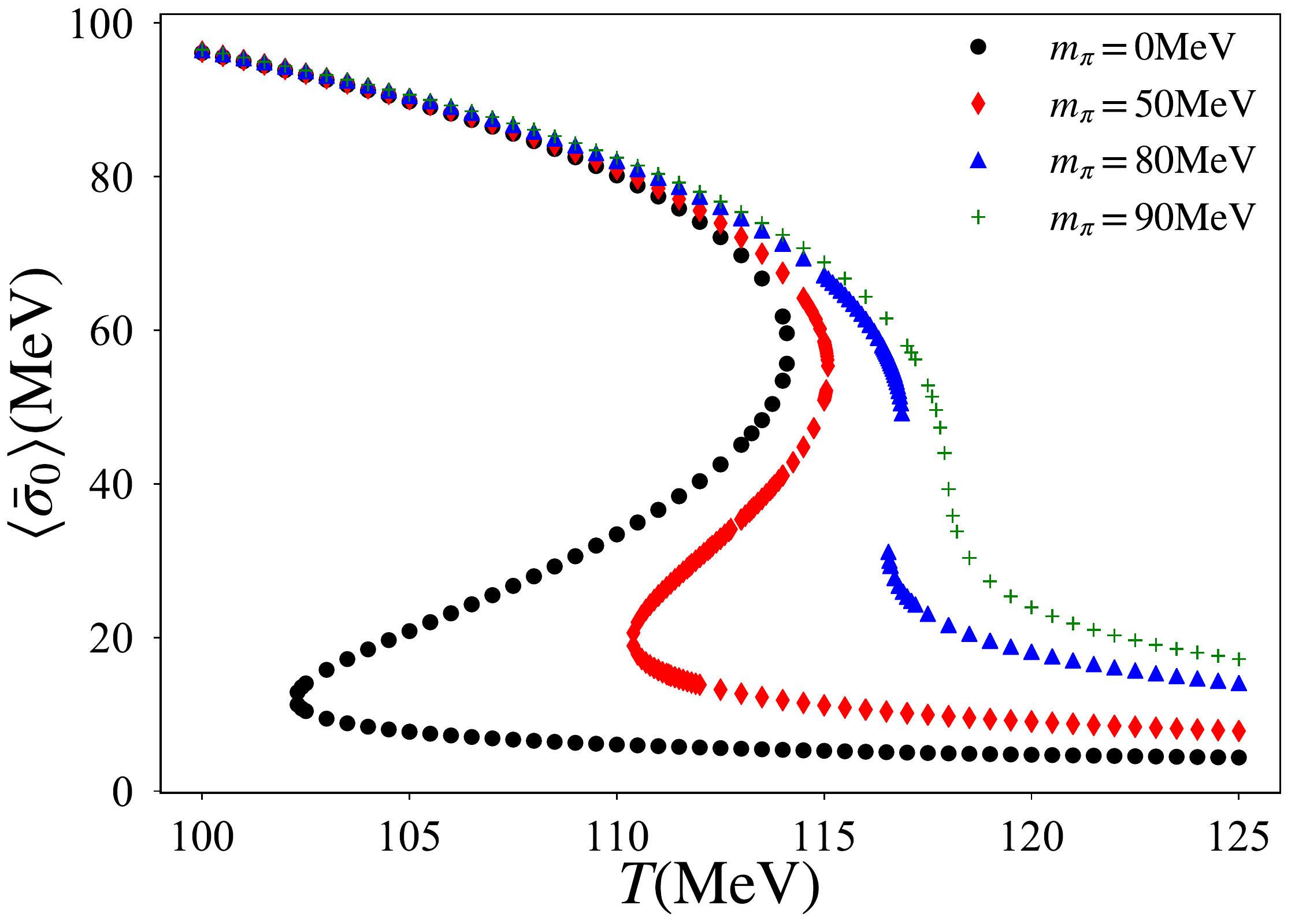}
    \subfigure{(a)}
 \end{center}
 \end{minipage}%
 \begin{minipage}{0.5\hsize}
 \begin{center}
   \includegraphics[width=7.0cm]{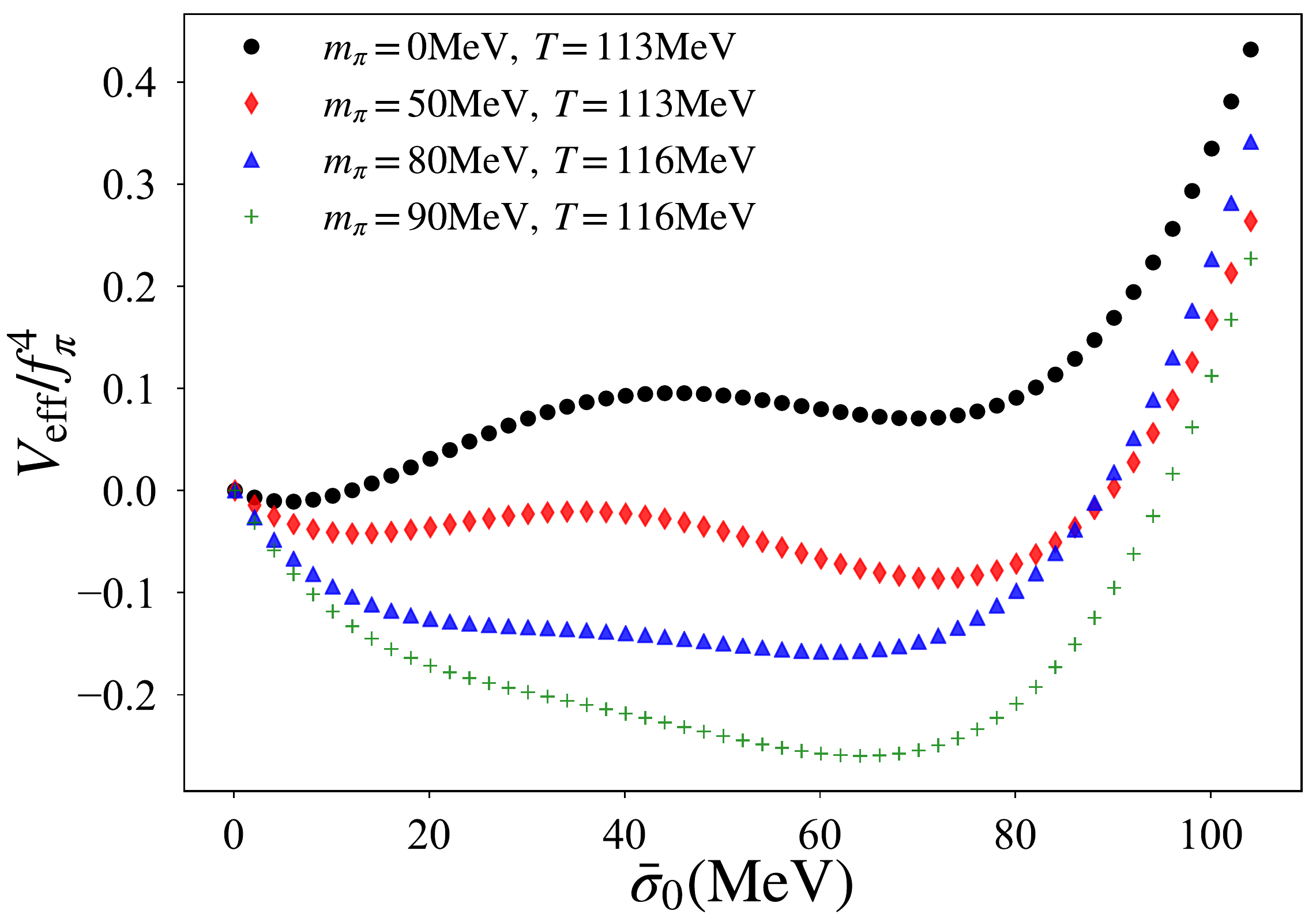}
    \subfigure{(b)}
  \end{center}
 \end{minipage}
 \end{tabular}
 \caption{
 Sensitivity of the pion mass to the chiral phase transition in the thermomagnetic QCD at $eB=f_\pi^2$, predicted 
 from the quark-meson model in the large $N_c$ limit, for the $SU(3)$ flavor 
 symmetric limit. 
 Panel (a): The $T$-evolution of the chiral order parameter $\bar{\sigma}_0$. Panel (b): The corresponding deformation of the effective potential, normalized in a way similar to the ones in Figs.~2 and~3 in the main text. 
 }  
 \label{nf3:mpi-changed}  
\end{figure}


Beyond the chiral limit, 
we have also investigated the pion-mass ($m_\pi$)
dependence on the chiral phase transition nature, 
in light of lattice simulations in the future. 
See Fig.~\ref{nf3:mpi-changed}, 
which corresponds to the $SU(3)$ flavor limit (the $N_f=3$ diagonal line in Fig.~1). 
Varying $m_\pi$ from zero, 
we observe a critical point, $m_\pi \simeq 80$ MeV, at which 
the transition nature changes from the first order 
to the second order, and will then be the crossover as gets closer to the physical pion mass.  
This places another critical endpoint 
in the extended Columbia plot, in  Fig.~\ref{fig:columbia_plot} 
(marked as the black circular blob).

The estimated value of the critical pion mass, 
$m_\pi \simeq 80$ MeV, 
cannot be so serious, because the present model should 
include the model-systematic uncertainty, which may roughly be 
30\% when the values of $T_c$ between ours and the lattice QCD result's are compared. See also footnote~\ref{foot:Tpc} in the main text.

Existence of this another critical endpoint seems to be 
trivial, because the chiral crossover for the pion mass around the physical point has already been established on the lattice QCD in a strong magnetic field. 
In this sense, we have just confirmed that the 
extrapolation to a weaker magnetic field regime 
works fine, continuously leading to the second order from the crossover. 
Though being such trivial, this endpoint would also be 
deserved to explore on the lattice QCD in the future. 
In contrast, the critical endpoint claimed in the main text (in the $m_s$ axis,  arising as the interplay along the massless two and three flavor limits) is nontrivial due to 
the emergence of the dramatic change of the transition nature into the crossover (for two-flavor) from the first order (three-flavor).

\end{document}